\documentclass[12pt,a4paper]{article}               
\usepackage{listings}
\usepackage{techrep_icg}
\usepackage{amsmath}
\usepackage{caption}
\usepackage{subcaption}
\begin{document}
\reportnr{ICG–CVARLab-TR–002}               
\title{Augmented Reality Oculus Rift} 
\subtitle{Developement Of An Augmented Reality Stereoscopic-Render-Engine To  Extend Reality With 3D Holograms} 
\repcity{Graz}            
\repdate{\today}          
\author{Markus H\"oll}
\author{Nikolaus Heran}
\author[ICG]{Vincent Lepetit}
\newcommand{\TUGn}{Graz University of Technology}
\address[ICG]{Inst. for Computer Graphics and Vision \\ \TUGn, Austria}
\contact{Markus H\"oll}
\contactemail{mhoell@student.tugraz.at}

\begin{abstract}
This paper covers the whole process of developing an Augmented Reality Stereoscopig Render Engine for the Oculus Rift. To capture the real world in form of a camera stream, two cameras with fish-eye lenses had to be installed on the Oculus Rift DK1 hardware. The idea was inspired by Steptoe \cite{steptoe2014presence}. After the introduction, a theoretical part covers all the most neccessary elements to achieve an AR System for the Oculus Rift, following the implementation part where the code from the AR Stereo Engine is explained in more detail. A short conclusion section shows some results, reflects some experiences and in the final chapter some future works will be discussed. The project can be accessed via the git repository \url{https://github.com/MaXvanHeLL/ARift.git}.
\end{abstract}

\tableofcontents
\addcontentsline{toc}{section}{Abstract}


\newpage

\section{Introduction}

Augmented Reality (AR) is the modern approach to create some kind of graphic holograms and place them into reality. Creating virtual worlds and objects is already possible since a long time using computer graphics but those virtual worlds generated with computer graphics were so far strictly separated from the real world, without any connection. Augmented Reality is now the key point to change this co-existence by merging both, the virtual world and the real world, together into a common visualisation. It is basically an interface to create holograms which we know already from science fiction movies or games. Using this technology, we are able to extend the space of the real world with digital information and visualisations of a virtual world.

Virtual Reality (VR) on the other hand, is a complete immersion into a virtual environment. It is achieved by blending out the real world around completely. There are some key factors which are very important to guarantee a deep immersion into a virtual world, like a stereo 3D view, to which humans are accustomed from biological eyes and also a proper head rotation. Such a virtual 3D view can be achieved with special hardware, the so called head-mounted-displays (HMDs). Representatives of these HMDs are for example the Oculus Rift, HTC Vive and Sony's VR.

There are already some AR prototype devices out there, like the Google Glass, Microsoft's Hololens and Smartphones running AR applications. The AR immersions are however limited due to the hardware construction. We created an AR System which makes use of the superior immersion of an VR HMD. The Oculus Rift DK1 served as the basic hardware for our project. Extending the Oculus Rift with two fish-eye cameras (IDS uEye UI-122-1LE-C-HQ) on the front plate gave us the possibility to extend the VR hardware to an AR device. 

In the theoretical section \ref{sec:theory}, we will have a look at the key elements of camera calibration with fish eye lenses to capture the real world, creating virtual holograms and merging them both together. Also, we will discuss some theoretical basics of an Oculus Rift integration into an existing graphics engine.

The third section \ref{sec:implementation} is treating the practical implementation of the AR stereo engine. We will discuss the implementation here in more detail, explaining some important code parts like the main function, camera capturing, calibration, render loop, stereo 3D rendering  and the Oculus Rift integration.

The modified AR hardware and some results are shown in chapter \ref{sec:conclusion}.

And lastly we will discuss some future related work in chapter \ref{sec:future} on which we will continue working on.


\section{Theory}\label{sec:theory}

 \subsection{Computer Vision -  Capturing the real world} \label{sec:virtual-world}

 \subsubsection{Camera Models, 3D-2D Mapping}

Capturing the real world 3D space coordinates onto an undistorted 2D image is called a 3D-2D mapping \cite{brauer2001new}.  In order to compute such an undistorted mapping, the cameras have to be calibrated previously to find according intrinsic camera parameters, which are also called the camera matrix of a pinhole camera model. There are several different camera models which are used to calibrate cameras. The pinhole camera model is the most traditional one, which assumes a perfect aperture. This means a proper camera calibration is essential to achieve good results by computing the undistorted 3D-2D mapping. 

 \subsubsection{Fish-Eye Lenses, Wide Field-Of-View}

\begin{figure}[htpb]
  \centering
  \includegraphics[width=0.3\textwidth]{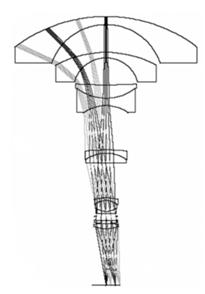}
  \caption{Camera model with fish eye lense. Image courtesy of "Scaramuzza et al. \cite{scaramuzza2006flexible}".  }
  \label{fig:fish-eye}
\end{figure}

Traditional camera systems have a very small field-of-view (FOV) about 45$^{\circ}$. These limitations are problematic for capturing the scene motion with a moving camera  \cite{gluckman1998ego}. Cameras with fish-eye lenses have the property of a very wide FOV. This property makes fish-eye lenses interesting in fields of photography and computer vision. Figure \ref{fig:fish-eye} illustrates such a fish-eye lense. 

Cameras with fish-eye lenses cause significantly higher image errors on the 3D-2D mapping due to the higher lense distortion. Especially on the edges the distortion is significantly higher than in the center. It is not suitable to calibrate cameras with such a high FOV using a traditional pinhole camera model. 

 \subsubsection{Omnidirectional Camera Calibration}

The omnidirectional camera model of Scaramuzza \cite{scaramuzza2006flexible} finds the relation between the 3D vector and a 2D pixel using a mirror or a fish-eye lens in combination with a camera.

\begin{figure}[htpb]
  \centering
  \includegraphics[width=0.5\textwidth]{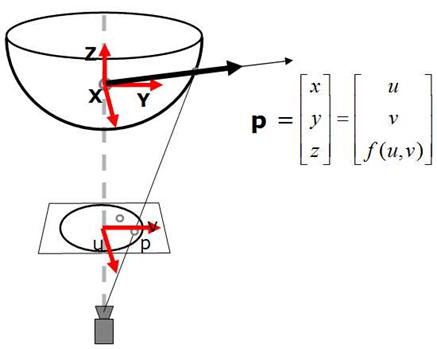}
  \caption{3D-2D coordinate mapping using the omnidirectional camera model. Image courtesy of "Scaramuzza et al. \cite{scaramuzza2006flexible}".  }
  \label{fig:omnimodel}
\end{figure}

3D-2D coordinate mapping from a 2D pixel and a 3D vector using Scaramuzza's omnidirectional camera model is illustrated in Figure \ref{fig:omnimodel}

The projection can be achieve by the  projection function $\textbf{f(p)}$ at equation \ref{eq:projectionf} which is a polynomial function. The coefficients of the projection function $\textbf{f(p)}$ are the calibration parameters and \textbf{p} describes the distance from the omnidirectional image center. The degree of polynom can be chosen, however, according to Scaramuzza he experienced best results with a polynom of 4.

 \begin{equation}
    f(\mathbf{p}) = a_0 + a_1p + a_2p^2 + a_3p^3 + a_4p^4 + ... + a_Np^N
    \label{eq:projectionf}
  \end{equation} 
 
\begin{figure}[htpb]
  \centering
  \includegraphics[width=1.0\textwidth]{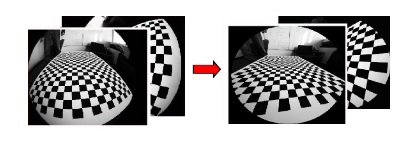}
  \caption{ \textbf{Left:} distorted camera image.  \textbf{Right:} undistorted camera image. Image taken from \cite{li2005non}.  }
  \label{fig:distortion}
\end{figure}

After finding the calibration parameters, the lense distortion can be corrected by finding image point corresponendences. The result is an undistorted image mapped from 3D to 2D, illustrated in Figure \ref{fig:distortion}

\newpage

\subsection{Computer Graphics - Creating virtual 3D holograms} \label{sec:virtual-world}

\subsubsection{Architecture}

\begin{figure}[htpb]
  \centering
  \includegraphics[width=0.48\textwidth]{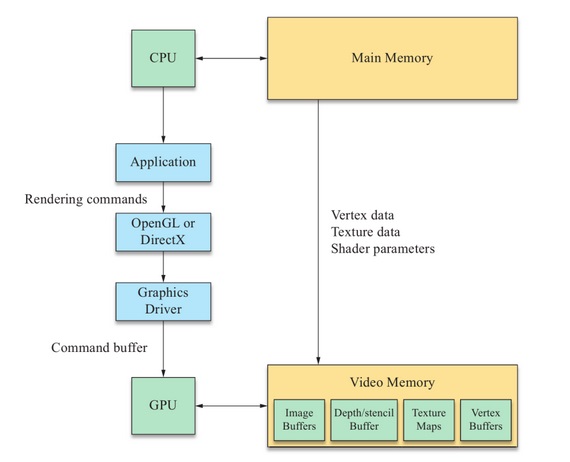}
  \caption{Architecture of a graphic interface. Image taken from \cite{lengyel2005mathematics}.  }
  \label{fig:architecture}
\end{figure}

Computer graphics is a key element to extend reality with holograms. Programming computer graphics is different, because the code is accessed by the GPU instead of the CPU which works in a different way. The main reason for this is because the GPU's architecture is totally different from CPU architecture. CPU cores are designed to run single threads as fast as possible, with a huge cache and smaller algorithmic logic unit (ALU). GPUs on the other hand are designed to run highly parallel. Therefore, a graphic API is needed to communicate with the GPU, for example DirectX, OpenGL or Vulkan. An abstract architecture is shown in Figure \ref{fig:architecture}

\subsubsection{Render Pipeline}

\begin{figure}[htpb]
  \centering
  \includegraphics[width=1.0\textwidth]{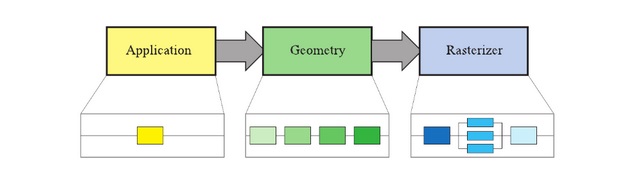}
  \caption{ Abstraction of a basic render pipeline. Image taken from \cite{akenine2008real}.  }
  \label{fig:pipeline}
\end{figure}

Programs running on a GPU are called shaders, which are written for example in HLSL when using DirectX API, or GLSL when using OpenGL API. Some years ago, the render pipeline was a fixed function pipeline, which means their funcitonality were implemented in hardware without customization. Since the invention of the unified shader model there are alot of stages that are individually programmable by developers using different kinds of shaders for different purposes. Each graphics engine has to have at least one vertex- and one fragment shader implemented. Modern graphic engines also have additional geometry- and tesselation shaders between the vertex- and the fragment shader. Figure \ref{fig:pipeline} shows a very rough abstraction of the render pipeline.

\subsubsection{Application}

\begin{figure}[htpb]
  \centering
  \includegraphics[width=0.6\textwidth]{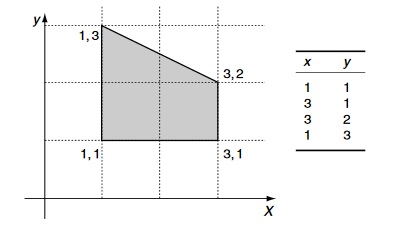}
  \caption{Simple 2D polygon defined by 4 different vertices in carthesian xy-plane. Image taken from \cite{vince2013mathematics}.  }
  \label{fig:polygon2D}
\end{figure}

The application stage takes care of user inputs and is executed on the CPU. Further on, the application stage feeds the geometry stage with geometric primitives (points, lines and triangles) like Akenine \cite{akenine2008real} points out. Virtual objects are constructed by defining vertices (points) within a space and computing polygons from them. Complex 3D models are designed previously using modeling programs like Blender, Maja or 3dsMax. The model files (.obj, .3ds, .fxb etc.) consisting of  the vertices are then loaded through the application stage.  Figure \ref{fig:polygon2D} shows how a simple 2D polygon would be defined on a carthesian xy-plane.

\subsubsection{Geometry}

\begin{figure}[htpb]
  \centering
  \includegraphics[width=1.0\textwidth]{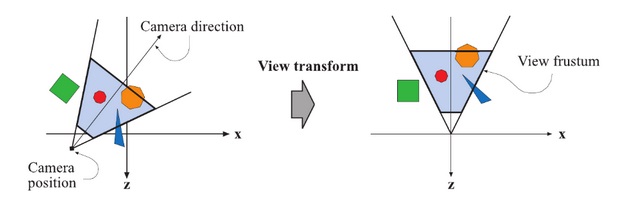}
  \caption{model-view transformation - placing models into the virtual world and transforming camera's position to the origin. Image taken from \cite{akenine2008real}.  }
  \label{fig:modelviewtransform2D}
\end{figure}

The programmable vertex shader is part of the geometry stage. In the geometry stage, some important computations are the model-view transform, plane projection and clipping. The model transform is placing the model's vertex coordinates into the virtual world $R\textsuperscript{3}$. The 3D scene is obsevered by a virtual camera which is placed also in the world with a specific translation and rotation. The view-transform places the camera to the origin and this transformation is further on applied to all of the rendered models. This is done because the view computation is much easier with this approach. The view transformation is illustrated in Figure \ref{fig:modelviewtransform2D}.

\begin{figure}[htpb]
  \centering
  \includegraphics[width=1.0\textwidth]{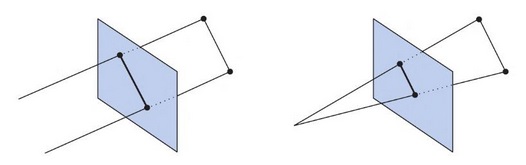}
  \caption{\textbf{Left:} orthographic projection.  \textbf{Right:} perspecitve projection. Image taken from \cite{akenine2008real}.  }
  \label{fig:projection}
\end{figure}

Projecting now the observed 3D scene onto a 2D plane is the next step. According to Akenine \cite{akenine2008real}, there are 2 commonly used projection methods depending on the viewing volume, namely a ortographic or a perspective projection. In our case, the viewing volume is a frustum which causes the effect that object which are farther away appear smaller. This can be done using perspecitve projection, which is shown in Figure  \ref{fig:projection}.

\newpage

This is the minimum functionality which a vertex shader has to do at least per model vertex. The whole process can be done by multiplying each model's homogenous vertex with 3 matrizes which are called the model matrix, view matrix and projection matrix shown in  Equation \ref{eq:projection}

\begin{equation}
  \mathbf{xy}_{\text{screen coordinates}} = \mathbf{Homogenous Vertex} * \mathbf{M}\mathbf{V}\mathbf{P}
  \label{eq:projection}
\end{equation}

Clipping is used to render only objects which are (partially) in the virtual camera's frustum to avoid lower performance. With all that, the most basic functionality of the geometry stage is roughly covered.


\subsubsection{Rasterizer}

\begin{figure}[htpb]
  \centering
  \includegraphics[width=0.3\textwidth]{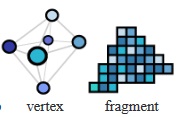}
  \caption{\textbf{Left:} Model in vertex stage.  \textbf{Right:} Model in rasterization stage. Image taken from \cite{proudfoot2001real}.  }
  \label{fig:fragment}
\end{figure}

The goal of this stage is to give the pixels their final color values. Post processing computation, texturie mapping, depth testing, illumination and alot more can be done with a fragment shader. One pixel can be covered by more than one object, for each of those fragments the computation takes place to color the pixel. Depth testing also takes place fragment-wise which is  computed using z-value comparison of each fragment's vertices. An illustration of the Polygon rasterization can be seen in Figure \ref{fig:fragment}.

\begin{figure}[htpb]
  \centering
  \includegraphics[width=0.8\textwidth]{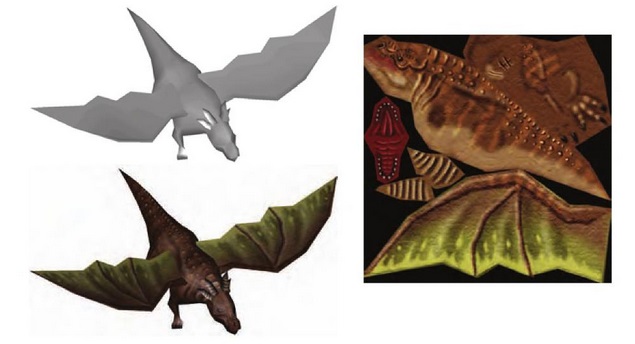}
  \caption{\textbf{Left:} 3D Model with- and without texture mapping.  \textbf{Right:} 2D texture. Image taken from \cite{akenine2008real}.  }
  \label{fig:dragon}
\end{figure}

The process of texturing gives 3D polygons a more realistic appearance. According texture coordinates can be read from each visible fragment to compute a texture mapping on 3D models. Those texture coordinates are commonly defined in the model files and defined by \textbf{u} and \textbf{v} coordinates in a 2D image space. The effect of texture mapping is shown in Figure  \ref{fig:dragon}.

The lighting is also computed withing the fragment shader, because it makes no sense calculating lighting beforehand at the vertex stage where lighting computation even would take place on vertices who are not even in the clipping space and therefore not visible for the observer. Instead, lighting is computed fragment wise. The Bidirectional-Reflectance-Distribution-Function (BRDF) of a simple lambertian shading with a perfectly diffuse reflector is given by

 \begin{equation}
    f_r = \dfrac{\mathbf{c}_d}{\pi}          
   \label{eq:brdf}
  \end{equation} 

The diffuse reflectance is given by $\textbf{c}_d$, meaning the fragment's color.  When the surface point is only illuminated by a single light source, the radiance  $\textbf{L}_0$ is then given by the Equation \ref{eq:illumination}.

 \begin{equation}
   \mathbf{ L}_0(\mathbf{n, s}) = \dfrac{\mathbf{c}_d}{\pi}{  \circ  \mathbf{E}_L max(\mathbf{n \cdot s, 0})}                             
   \label{eq:illumination}
  \end{equation} 

\begin{figure}[htpb]
  \centering
  \includegraphics[width=0.36\textwidth]{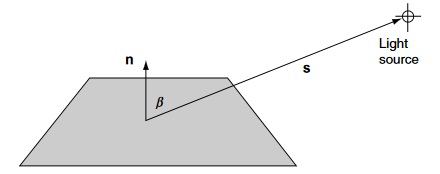}
  \caption{single point light illuminating a surface. Image taken from \cite{vince2013mathematics}.  }
  \label{fig:light}
\end{figure}

 $\textbf{n}$ is the surface normal vector of the fragment,  $\textbf{s}$ is the the light direction from the surface point to the light source and  $\textbf{E}_L$ is the irradiance from the light source, meaning the light color. Figure \ref{fig:light} illustrates a surface illuminated by a single point light.

\newpage

\subsection{Augmented Reality - Merging both worlds together}

With the real world captured and the virtual world fully created, it is now possible to merge both information streams together into one visualization to achieve a first basic AR system. It is therefore neccessary to render the camera stream into a dynamic 2D texture, which gets updated on each render cycle. The camera stream has to be rendered orthogonal to the screen. 

To avoid a depth overlap, the camera stream is rendered without z-buffering, which means it will be treated as a background plane from a computer graphics point of view. With this, it is warranted that the created holograms will be rendered visible for the observer without colliding with the camera stream.

\begin{figure}
    \centering
    \begin{subfigure}[b]{0.31\textwidth}
        \includegraphics[width=\textwidth]{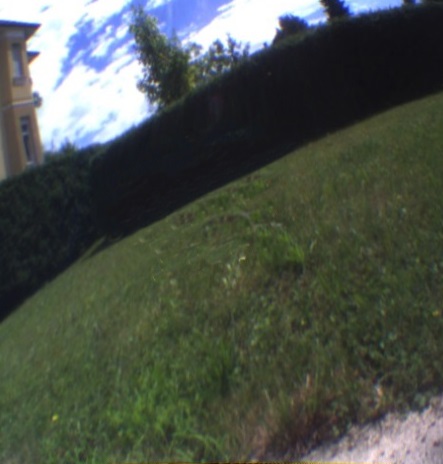}
        \caption{camera stream}
        \label{fig:reality}
    \end{subfigure}
    ~ 
    \begin{subfigure}[b]{0.31\textwidth}
        \includegraphics[width=\textwidth]{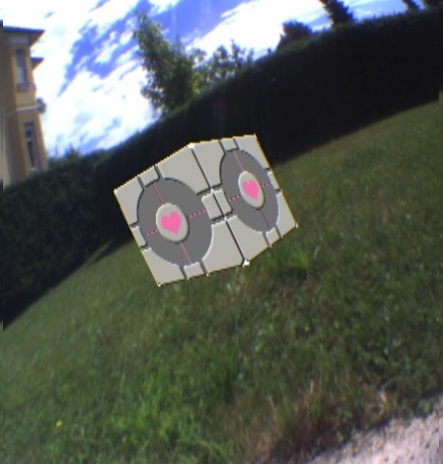}
        \caption{Augmented}
        \label{fig:augmented}
    \end{subfigure}
    ~ 
    \begin{subfigure}[b]{0.31\textwidth}
        \includegraphics[width=\textwidth]{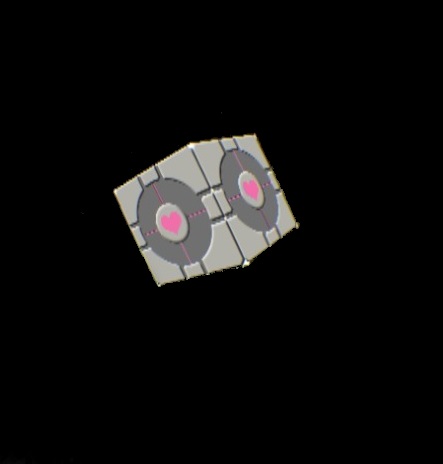}
        \caption{3D hologram}
        \label{fig:hologram}
    \end{subfigure}
    \caption{Merging the holograms}\label{fig:worlds}
\end{figure}

The virtual world however, is just a space in $R\textsuperscript{3}$ without a visible ground, sky or anything like that. It is just a space to place the virtual 3D models in. The merging process of a textured 3D cube hologram is shown in  Figure \ref{fig:worlds}.

\subsection{Virtual Reality Immersion - Stereoscopic Rendering}

\subsubsection{3D View}

\begin{figure}[htpb]
  \centering
  \includegraphics[width=0.6\textwidth]{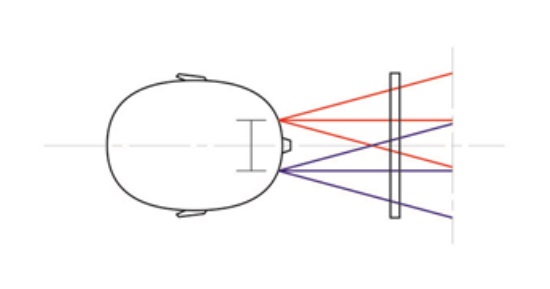}
  \caption{HMD's eye view cones. Image taken from \cite{oculusDeveloperGuideV006}.  }
  \label{fig:rift_eyeviews}
\end{figure}

With HMDs, it is possible to immerse fully into virtual worlds like computer games. HMDs are designed to give a stereoscopic view of the virtual scene like we are used to with human eyes. This understanding of depth is achieved by observing a scene from 2 different cameras, normally our eyes, which are slightly translated along the x-axis. The human eyes have a interpupillary distance (IPD) of approximately 65 mm as stated here \cite{oculusDeveloperGuideV006}. The Oculus Rift is designed in a way that the left eye sees the left half of the intern screen and the right eye sees the right half of the intern screen as illustrated in Figure \ref{fig:rift_eyeviews}.

Achieving a natural 3D view from the real world is simply achieved by using 2 stereo cameras translated by a IPD of about 65 mm aswell. The camera stream as a 2D texture can later on be translated and adjusted accordingly to achieve a correct 3D view of the real world.

To integrate a HMD to the AR engine, it is therefore needed to render the whole AR scene twice, alternating the real world camera stream aswell as translating the virtual camera according to the IPD of human eyes and rendering the virtual scene from different viewpoints with that. 

\subsubsection{Post-Processing, Image Distortion}

\begin{figure}[htpb]
  \centering
  \includegraphics[width=0.5\textwidth]{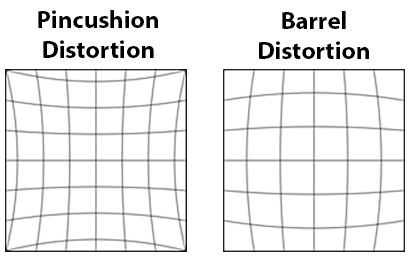}
  \caption{barrel distortion to counteract the lense-based pincushion distortion of the HMD. Image taken from \cite{oculusDeveloperGuideV006}.  }
  \label{fig:barreldistortion}
\end{figure}

However, since the Oculus Rift enhanced the virtual reality immersion through a very wide FOV achieved by the lenses, the original images would show a pincushion distortion as pointet out by Oculus \cite{oculusDeveloperGuideV006}. To counteract that, the images have to be post-processed by applying a barrel distortion shown in Figure  \ref{fig:barreldistortion}.

\subsubsection{Stereoscopic Projection Transformation}

\begin{figure}[htpb]
  \centering
  \includegraphics[width=0.65\textwidth]{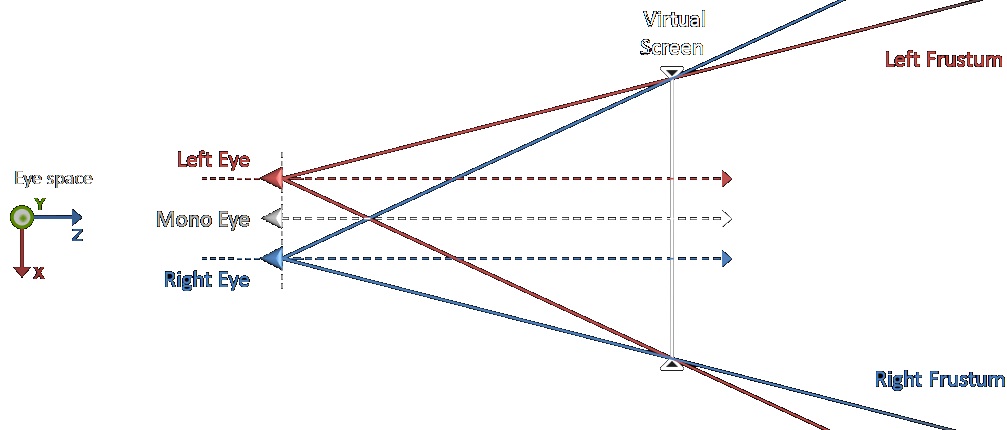}
  \caption{stereo projection transformation for stereoscopig 3D projection. Image taken from \cite{nvidia2011stereo}.  }
  \label{fig:stereoprojectiontransformation}
\end{figure}

The perviously used projection matrix based on a perspective projection can no longer be used in stereoscopic rendering. Therefore, it is not sufficient to only translate the virtual cameras along the \textbf{x}-axis, because the cameras would no longer project at the same plane. Thus the projection matrix has to be modified to compute a stereo projection transformation as pointet out by Nvidia \cite{nvidia2011stereo}. Stereo projection transformation is shown in Figure \ref{fig:stereoprojectiontransformation}.

According to Nvidia \cite{nvidia2011stereo}, a general computation of the stereo projection transformation can be achieved by modifying the normal perspective projection matrix.

Left handed row major matrix (D3D9 style):
\[
Projection_{stereo} =
 \begin{bmatrix}
  \begin{smallmatrix}
    p11 & 0 &0  &  0  \\
    0 & p22 & p32 & 0  \\
    p13 + \textbf{side} * \textbf{separation}   & 0 & p33 & 1  \\
    -\textbf{side} * \textbf{separation} * \textbf{convergence} & 0 & p34 & 0
  \end{smallmatrix}
\end{bmatrix}
\]

Right handed column major matrix (OpenGL style):
\[
Projection_{stereo} =
 \begin{bmatrix}
  \begin{smallmatrix}
    p11 & 0 & p13 - \textbf{side} * \textbf{separation}  & -\textbf{side} * \textbf{separation} * \textbf{convergence}  \\
    0 & p22 & p23 & 0  \\
    0  & 0 & p33 & p34  \\
    0 & 0 & -1 & 0
  \end{smallmatrix}
\end{bmatrix}
\]

where \textbf{side} is -1 for left and +1 for right, \textbf{pij} are coefficients of the mono perspective projection, \textbf{convergence} is the plane where left and right frustums intersect and \textbf{separation} is the normalized version of interaxial by the virtual screen width.

\subsubsection{Head Rotation, Hardware Sensor}

\begin{figure}[htpb]
  \centering
  \includegraphics[width=0.5\textwidth]{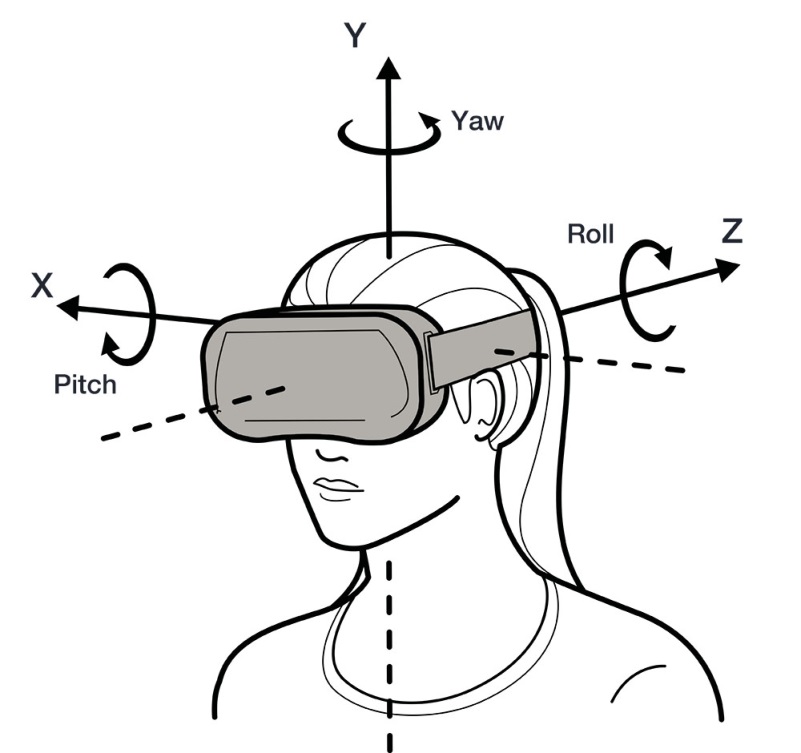}
  \caption{Head rotation using gyro sensor from the Oculus Rift. Image taken from \cite{oculusDeveloperGuideV006}.  }
  \label{fig:rift_rotation}
\end{figure}

Further on, the Oculus Rift's hardware integrated gyro sensor can be used to gain information about the human's head rotation (yaw, pitch and roll). They can now be used to synchronize the head rotation with the virtual mono camera to apply the same kind of view rotation observing the 3D holograms. The head rotation is illustrated in Figure \ref{fig:rift_rotation}.

\paragraph{Stereoscopic AR Rendering:}
  \begin{enumerate}
   \item receive head rotation from gyro sensor integrated in the Oculus Rift hardware and apply rotation to virtual mono camera.
   \item Translate virtual mono camera about ~32 mm along the  \textbf{ negative x}-axis and render AR scene with \textbf{ left} camera stream into 2D texture.
   \item Translate virtual mono camera about ~32 mm along the  \textbf{ positive x}-axis and render AR scene with \textbf{ right} camera stream into 2D texture.
    \item Set the Swapchain's RenderTargetView now finally to the \textbf{Backbuffer}, to present the upcoming stereo image on the application's screen.
    \item supply the Oculus SDK with the 2 stereo rendered 2D textures and initiate the post-processing \textbf{barrel distortion}, provieded by the Oculus SDK.
  \end{enumerate}

\newpage


\lstset{language=C++,
       numbers=left,
        basicstyle=\ttfamily,
        keywordstyle=\color{blue}\ttfamily,
        commentstyle=\color{green}\ttfamily,
        frame=single,
        rulecolor=\color{black},    
        breaklines=true,
}

\section{Implementation}\label{sec:implementation}

The code of the project is located at a public GIT repository which can be accessed via \url{https://github.com/MaXvanHeLL/ARift.git}.

\subsection{Hardware Modification}

To develop an AR system, it was of course neccessary to add both of the IDS uEye UI-122-1LE-C-HQ cameras in a way to capture stereo images. The cameras are bolted onto a plastic glass which is wired ontop of the front plate of the Oculus Rift DK1. The modified hardware is shown below.

 \begin{figure}[!htb]
    \centering
    \label{fig:hardware}
    \begin{subfigure}{0.45\textwidth}
        \includegraphics[width=\textwidth]{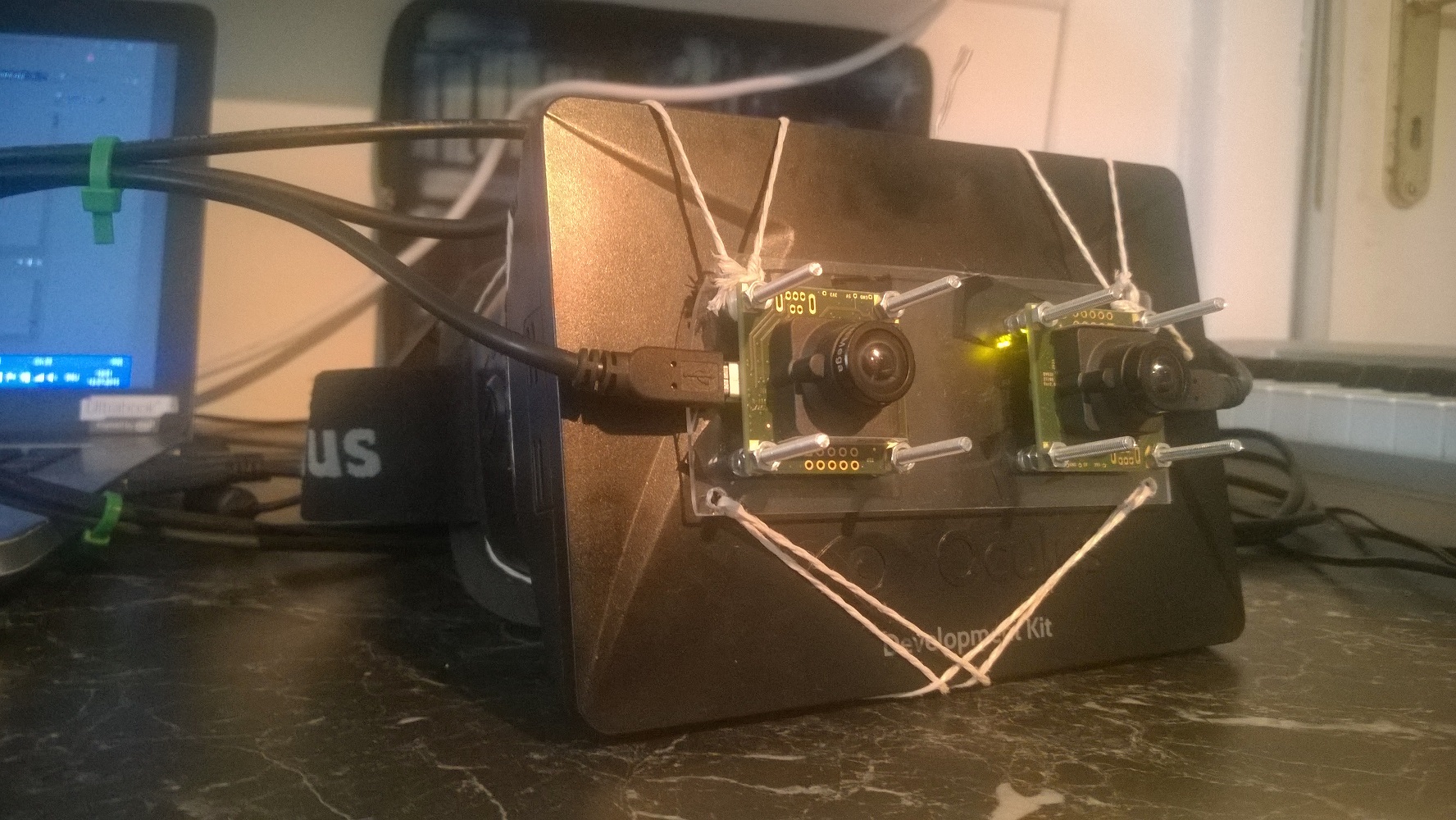}
        \label{fig:hardware1}
    \end{subfigure}
    ~ 
    \begin{subfigure}{0.45\textwidth}
        \includegraphics[width=\textwidth]{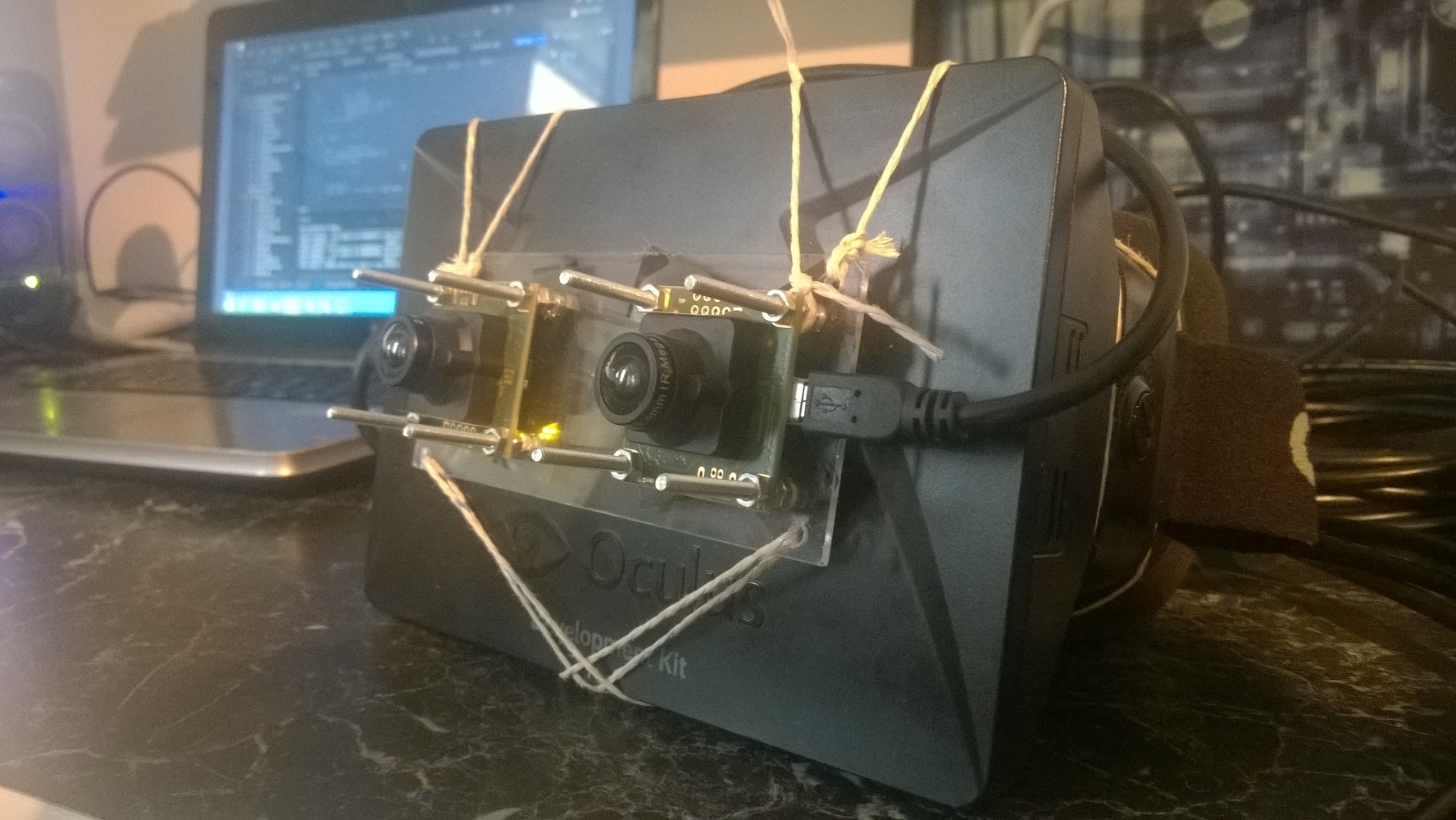}
        \label{fig:hardware2}
    \end{subfigure}
    \caption{Hardware Modifications on Oculus Rift DK1}
\end{figure}
\label{fig:hardware}

\subsection{OCamCalib Matlab Toolbox}

Using the OCamCalib Omnidirectional Camera Calibration Toolbox for Matlab, developed by Scaramuzza \cite{scaramuzza2006toolbox} it was possible to find intrinsic camera parameters for both of the IDS uEye UI-122-1LE-C-HQ cameras which were used in this project.

\begin{figure}[H]
    \centering
    \begin{subfigure}[b]{0.30\textwidth}
        \includegraphics[width=\textwidth]{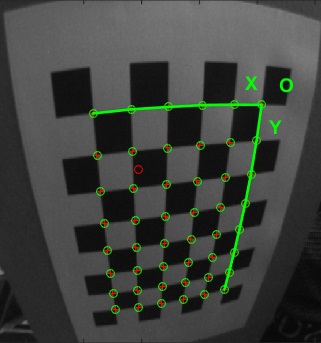}
        \label{fig:corners1}
    \end{subfigure}
    ~ 
    \begin{subfigure}[b]{0.30\textwidth}
        \includegraphics[width=\textwidth]{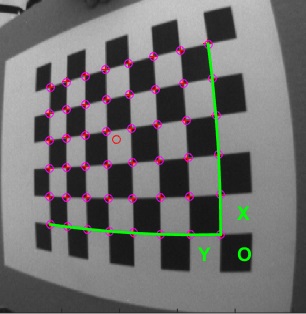}
        \label{fig:corners2}
    \end{subfigure}
    \caption{Calibration -Automatic Corner Detection}\label{fig:corners}
\end{figure}

The toolbox implemented also an automative corner detection on a checker board sample, shown in Figure \ref{fig:corners}

A visualization integrated in the toolbox shows the extrinsic camera parameters due to the samples images that have been taken. The illustration below shows the OcamCalib visualization.

 \begin{figure}[htb]
    \centering
     \includegraphics[width=0.6\textwidth]{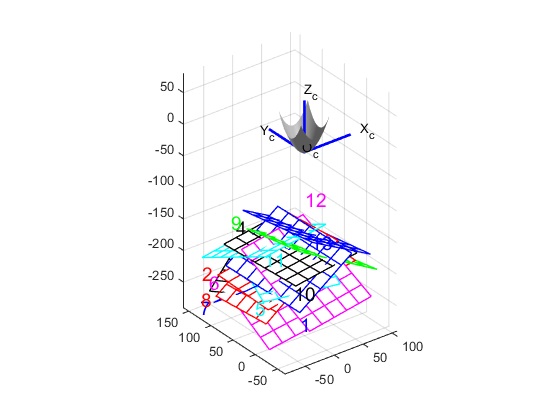}
     \label{fig:extrinsic}
    \caption{Calibration - Extrinsic Camera Parameters}
\end{figure}

\subsection{Engine Design}

\begin{figure}[!htb]
  \centering
  \includegraphics[width=1.0\textwidth]{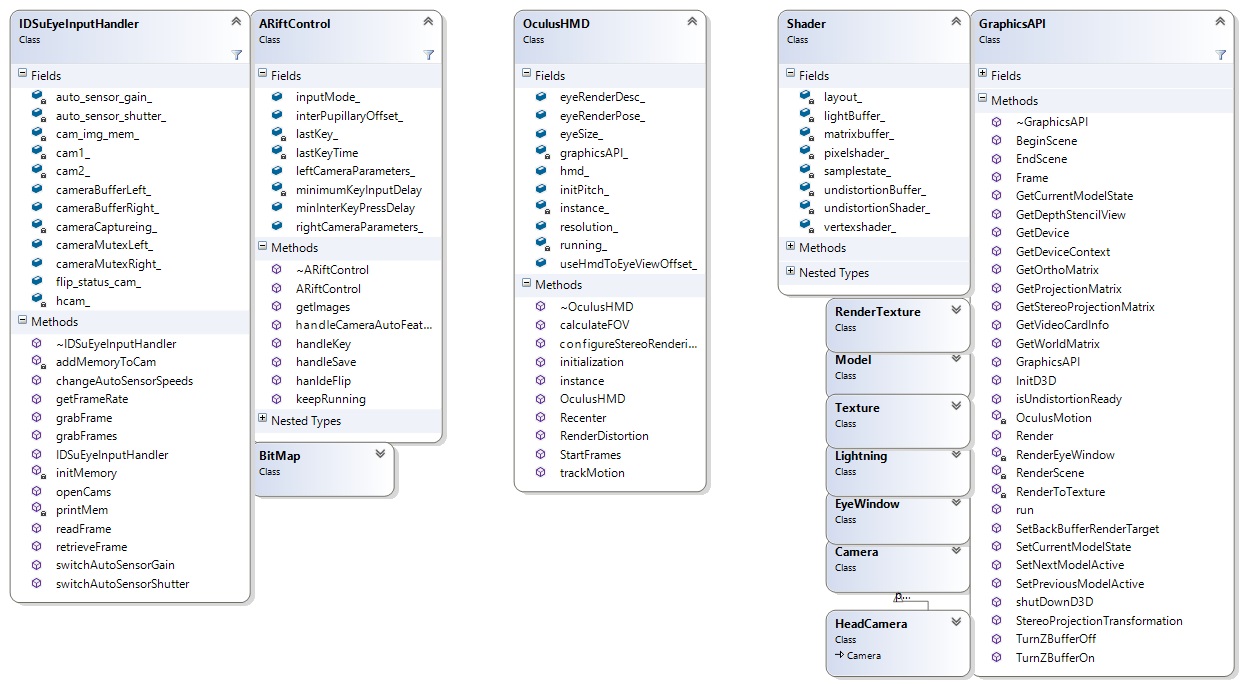}
  \caption{AR stereo engine code design. Image property of Markus H\"oll.  }
  \label{fig:program}
\end{figure}

The AR Stereo Engine has been written in DirectX11 using C++. It has basically to handle 3 major assignments, on the one hand the image capturing, creation the virtual scene and lastly melting them both together. All the classes are illustrated in Figure \ref{fig:program}. The left classes concern about the camera stream, on the right side are the graphics related classes and in the middle is the Oculus integration class. The scale factor of the class is standing for the importance of the class to give a better picture of the engine's design. The core of the whole system is, however, the \textbf{GraphicsAPI}.

On the left side there is the ARiftControl class, which takes care of input handling and getting images from camera image buffers. All the low-level image capturing from the cameras had to be written manually, because due to the bandwith  limitations of USB 2.0 cameras, it was not possible to stream both cameras parallel using OpenCV over the same HUB.  So it was neccessary to implement the image capturing manually, and thats what the IDSuEyeInputHandler is used for. The class directly uses IDSuEye Driver functions to initialize the cameras, allocating memory to them and retrieving image frames.

On the right side, there are the graphic related classes. The GraphicsAPI class is the core of the whole AR Engine which means it is communicating with all other graphics classes.

The OculusHMD static class is basically the interface to the Oculus Rift. With that, the HMD's initialization, post-processing rendering and head rotation synchronization can be done.

EyeWindow is only used as an additional functionality when the user is rendering with the stereo engine but without an HMD device. It takes care of placing the stereo images besides each other on the screen itself without the post-processing barrel distortion.

RenderTexture saves the rendered scene into two 2D texture, which later on gets supplied to the OculusHMD instance.

Model represents the 3D computer graphic objects, which means the actual holograms.

Each Model has its own Texture object which can be loaded to compute a texture mapping on them.

The BitMap is needed for rendering the camera stream orthogonal to the screen which is handled of course differently to normal 2D textures here.

Camera class is  like a normal virtual camera.

HeadCamera is further used to simulate the head rotations of the real head into the virtual camera. All the 3D view rendering is done with that.

Lighting handles the virtual illumination, which means light direction, color of the light aswell as ambient lighting.

The Shader serves as an interface to the GPU. It holds all data neccessary for the shader programs and inits the rendering.

Since explaining the whole engine is very complex and huge, only a rough explanation about the initialization can be gathered in chapter \ref{sec:init}. Further on, the basic stereo render loop is explained in chapter \ref{sec:stereorenderloop}. If more code detailes are desired, chapter \ref{sec:codedetails} gives a deeper look into the actual implementation. Though, there are only some code details shown which are also very abstracted for the sake of readability. Section \ref{sec:config} lists some program configurations which can be enabled or disabled and chapter \ref{sec:library} shows the used librarys.

\subsection{Initialization} \label{sec:init}

The main function basically does an initialization of the GraphicsAPI, which is the core of the whole engine. Also the ARiftControl instance gets initialized here, but only if an AR HMD device is plugged in to the computer. This can be controlled via the define \texttt{AR\_HMD\_ENABLED}.

The ARiftControl instance is allocated and initiated by default. Details about ARiftControl can be taken from \ref{sec:ARiftControl}.

Since the Constructor of GraphicsAPI is called in \textbf{line 4}, all the members which are held inside there, like an instance of ARiftControl (to retrieve a camera image later on as a 2D texture), both of the DirectX11 interfaces (RenderDevice, RenderContext), HeadCamera, BitMap, Model Vector, Shader, Lighting and so on, get initialized with 0 in the beginning. They will be assigned correctly later on.

After that, ARiftControl will be initialized properly in \textbf{line 9}, which is explained in more detail at \ref{sec:ARiftControl}.

\begin{lstlisting}
int main(int, char**)
{
  // DirectX Graphics and OculusHMD
  dx11 = new GraphicsAPI();
  HANDLE handle_render_thread = 0;
  ARiftControl cont;

  if (AR_HMD_ENABLED)
    cont.init(dx11);

   // Activate the Graphics (DirectX11) Thread
   handle_render_thread = CreateThread(NULL, 0,
    directXHandling, &cont, 0, NULL);
}
\end{lstlisting}

The project was always considered to be multi-threaded since it was clear from the start that in future work it will be developed further. This is the reason why the engine is running in another thread besides of the main thread which makes it easily extendable further on and that is what \textbf{line 12} is doing, starting the projects actual core thread.

\begin{lstlisting}
DWORD WINAPI directXHandling(LPVOID lpArg)
{
  ARiftControl* arift_c = (ARiftControl*)lpArg;

  if (AR_HMD_ENABLED)
  {
    // install the Oculus Rift 
    OculusHMD::initialization(dx11);
    OculusHMD::instance()->calculateFOV();
  }

  dx11->window_class_.cbSize = sizeof(WNDCLASSEX);
  dx11->window_class_.style = CS_HREDRAW | CS_VREDRAW;
  dx11->window_class_.lpfnWndProc = WindowProc;
  dx11->window_class_.hInstance = dx11->hinstance_;
  dx11->window_class_.hCursor = LoadCursor(NULL, IDC_ARROW);
  dx11->window_class_.hbrBackground = (HBRUSH)COLOR_WINDOW;
  dx11->window_class_.lpszClassName = dx11->applicationName_;

  RegisterClassEx(&dx11->window_class_);

  // application window
  dx11->window_ = CreateWindowEx(NULL, dx11->applicationName_, L"DirectX Render Scene",   
     WS_OVERLAPPEDWINDOW,  0, 0, RIFT_RESOLUTION_WIDTH, RIFT_RESOLUTION_HEIGHT,   
    NULL,  NULL, dx11->hinstance_, NULL);                    

  dx11->InitD3D(RIFT_RESOLUTION_WIDTH, RIFT_RESOLUTION_HEIGHT, VSYNC_ENABLED, dx11->window_, 
    FULL_SCREEN, SCREEN_DEPTH, SCREEN_NEAR, arift_c);
  ShowWindow(dx11->window_, SW_SHOW); 	
  SetFocus(dx11->window_);

   if (AR_HMD_ENABLED)
     OculusHMD::instance()->configureStereoRendering();		

  // msg loop
  while (TRUE)
  {
     // new msg?
     while (PeekMessage(&msg, NULL, 0, 0, PM_REMOVE))
     {
        TranslateMessage(&msg);
        DispatchMessage(&msg);
     }
    // quit
    if (msg.message == WM_QUIT)
      break;

    if (msg.message == WM_CHAR)
    {
      //trigger ARiftControl inputhandling
      arift_c->handleKey((char)msg.wParam);
    }
    // Run "engine" code here
    // -----------------------
    arift_c->camInput_->grabFrames();
    frame_return = dx11->Frame();
    // -----------------------
  }
  return msg.wParam;
}
\end{lstlisting}
\textbf{Line 8} is initializing the static Oculus Rift if the \texttt{AR\_HMD\_ENABLED} define is enabled and \textbf{Line 9} calculates the FOV according to the hardware specification. The Oculus Rift configuration is explained in more detail at \ref{sec:OculusHMD}.

Further on, the applications main window is initialized where the DirectX scene will be rendered to. In \textbf{Line 27} the GraphicsAPI gets initialized properly and sets up all the graphic related parts. It is recommended to look into the GraphicsAPI at section \ref{sec:GraphicsAPI} for more details. \textbf{Line 33} configures some parameters for the HMD stereo rendering.

The program enters then the main loop which is requesting input messages and iterating the render loop. If a msg which is not the WM QUIT message is triggered in form of a character, the input handling from \textbf{ARiftContol} \ref{sec:ARiftControl} handles the according input.

After that, the new camera frames get captured with \textbf{Line 55} and further on the next stereo render starts. Since the render loop is the core of the whole engine, it is explained
in chapter \ref{sec:stereorenderloop} in more detail.

\subsection{Stereo Render Loop}\label{sec:stereorenderloop}

\begin{lstlisting}
bool GraphicsAPI::Frame()
{
  // request head rotation from the Oculus Rift hardware
  OculusMotion();

  headCamera_->SetPosition(ariftcontrol_->cameraPositionX_, 
    ariftcontrol_->cameraPositionY_,  ariftcontrol_->cameraPositionZ_);

  // Render the virtual scene.
  result = Render();

  return result;
}
\end{lstlisting}
First of all, the Oculus' current head rotation is captured using the gyro sensor. Details about that can be seen in chapter \ref{sec:OculusHMD}. The virtual camera's rotation is synchronized with the oculus rotation here.
Further on, the \textbf{Render()} method is called.

\begin{lstlisting}
bool GraphicsAPI::Render()
{

  if (HMD_DISTORTION && AR_HMD_ENABLED)
    OculusHMD::instance()->StartFrames();

  // [Left Eye] first pass of our rendering. 
  result = RenderToTexture(renderTextureLeft_, 1);

  BeginScene(0.0f, 0.0f, 0.0f, 1.0f);

  if (!HMD_DISTORTION)
  {
    TurnZBufferOff();
    RenderEyeWindow(eyeWindowLeft_, renderTextureLeft_);
    TurnZBufferOn();
  }

  // [Right Eye] second pass of our rendering
  result = RenderToTexture(renderTextureRight_,2);

  if (!HMD_DISTORTION)
  {
    TurnZBufferOff();
    RenderEyeWindow(eyeWindowRight_, renderTextureRight_);
    TurnZBufferOn();
  }

  if (HMD_DISTORTION && AR_HMD_ENABLED)
    OculusHMD::instance()->RenderDistortion();
  else
    EndScene();

  return result;
}
\end{lstlisting}
In \textbf{Line 8} the scene gets rendered the first time where the virtual HeadCamera will be translated about ~32 mm along the negative x-axis. The render target will be set to the left \textbf{RenderTexture} instead of the backbuffer.
\textbf{Line 14} will trigger only when the post-processing barrel distortion from the Oculus SDK is not desired, therefore the rendering works differently because presenting both of the eye images is handled by the engine itself, without barrel distortion of course.
To also render the scene from the viewpoint of the right eye, \textbf{Line 20} calls the second renderpass where the virtual camera will be translated by ~32 mm along the positive x-axis and the rendered image will be store in a \textbf{RenderTexture} for the right eye.
Presenting both stereo images now depends on the chosen engine configuration. Only if \texttt{AR\_HMD\_ENABLED} and \texttt{HMD\_DISTORTION} are set to 1, the last render pass is handled by the Oculus SDK and post-processing barrel distortion is applied. If the engine is configured differently, \textbf{EndScene()} is called instead and the stereo image rendering is handled manually.

\subsection{Code Details} \label{sec:codedetails}

\subsubsection{GraphicsAPI}\label{sec:GraphicsAPI}

\begin{lstlisting}
void GraphicsAPI::OculusMotion()
{
  float oculusMotionX, oculusMotionY, oculusMotionZ;
  OculusHMD::instance()->trackMotion(oculusMotionY, oculusMotionX, oculusMotionZ);

  headCamera_->SetRotation(-oculusMotionX, -oculusMotionY, oculusMotionZ);
}
\end{lstlisting}
Receiving the head rotation using the HMD's gyro sensor and set it as the \textbf{HeadCamera}'s rotation aswell. Details about the implementation of the Oculus device can be taken from' chapter \ref{sec:OculusHMD}.

\begin{lstlisting}
bool GraphicsAPI::RenderToTexture(RenderTexture* renderTexture, int cam_id)
{
  // set render target to RenderTexture
  renderTexture->SetRenderTarget(devicecontext_, GetDepthStencilView());
  // clear the buffer
  renderTexture->ClearRenderTarget(devicecontext_, GetDepthStencilView(), 0.0f, 0.0f, 1.0f, 1.0f);

  // render scene into RenderTexture
  result = RenderScene(cam_id);

  // set next rendertarget  to the BackBuffer
  SetBackBufferRenderTarget();

  return result;
}
\end{lstlisting}
This method sets the next render target to a given \textbf{RenderTexture}, clears the buffer and renders the scene into it. Afterwards, the next rendertarget is set back to the BackBuffer again.

\begin{lstlisting}
bool GraphicsAPI::RenderScene(int cam_id)
{
  GetWorldMatrix(worldMatrix);
  GetOrthoMatrix(orthoMatrix);

  // ********** || 2D RENDERING || ***********
  TurnZBufferOff();
  result = shader_->Render(devicecontext_, bitmap_->GetIndexCount(), worldMatrix, cameraStreamMatrix, orthoMatrix,
    bitmap_->GetTexture(), undistBuffer, illumination_->GetDirection(), illumination_->GetDiffuseColor(),
  illumination_->GetAmbientColor());
  TurnZBufferOn();
  
  // ********** || 3D RENDERING || ***********
  // set head center to eye center offset
  headCamera_->headToEyeOffset_.position = ariftcontrol_->headToEyeOffset_;
  // Generate the view matrix
  headCamera_->RenderEye(cam_id == 1);
  // apply stereo projection transformation
  StereoProjectionTransformation(cam_id);
  // render all 3D models
  for (std::vector<Model*>::iterator model = models_.begin(); model != models_.end(); model++, i++)
  {
    result = shader_->Render(devicecontext_, (*model)->GetIndexCount(), worldMatrix, viewMatrix, stereoProjectionMatrix,
			model_tex, illumination_->GetDirection(), illumination_->GetDiffuseColor(), illumination_->GetAmbientColor());
  }
}
\end{lstlisting}
The method renders first the 2D bitmap with the camera as an image to the screen. Further, all the matrix computations are done and all 3D models are rendered here as well.

\begin{lstlisting}
void GraphicsAPI::StereoProjectionTransformation(int camID)
{
  Matrix4f proj = ovrMatrix4f_Projection(OculusHMD::instance()->eyeRenderDesc_[camID-1].Fov, screenNear_, screenDepth_, false);
  stereoprojectionmatrix_._11 = proj.M[0][0];
  stereoprojectionmatrix_._21 = proj.M[0][1];
  stereoprojectionmatrix_._31 = proj.M[0][2];
  stereoprojectionmatrix_._41 = proj.M[0][3];
  [...]
}
\end{lstlisting}
Due to stereo rendering, it is neccessary to modify the normal perspective projection matrix by applying a stereoscopic projection transformation. To achieve the correct modification for the Oculus hardware, the Oculus SDK is used here co compute the stereo projection transformation and copy to the intern stereo projection matrix for rendering.

\subsubsection{ARiftControl}\label{sec:ARiftControl}

\begin{lstlisting}
ARiftControl::ARiftControl()
{
  rightCameraParameters_.Nxc = 79.4f;
  rightCameraParameters_.Nyc = 67.2f;
  rightCameraParameters_.z = -177.0f;

  rightCameraParameters_.p9 = 0.0f;
  [...]
  rightCameraParameters_.p0 = 712.870100f;
  rightCameraParameters_.c = 1.000052f;
  rightCameraParameters_.d = 0.000060f;
  rightCameraParameters_.e = -0.000032f;
  rightCameraParameters_.xc = 236.646089f;
  rightCameraParameters_.yc = 394.135741f;
  rightCameraParameters_.width = 752.0f;
  rightCameraParameters_.height = 480.0f;
  
  // same for leftCameraParameters
 [...]
}
\end{lstlisting}
The constructor's assignment is primary to set all the camera parameters gained from the camera calibration for the left- and the right camera. 

\begin{lstlisting}
void ARiftControl::handleKey(char key)
{
  lastKeyTime = std::chrono::system_clock::now();
  switch (inputMode_)
  {
    case InputMode::DEFAULT: {...}
    case InputMode::MODEL: {...}
    case InputMode::WORLD: {...}
    case InputMode::CAMERA: {...}
    case InputMode::MOVEMENT: {...}
  }
}
\end{lstlisting}
The input handling is based on different modes like MODEL, CAMERA, DEFAULT, and so on. During DEFAULT mode, all actions go to the camera stream like translating both cameras or zooming in- and out of the camera stream. MODEL mode is used to manipulate the virtual objects and translating them in the virtual 3D space, CAMERA is doing the same but with the virtual mono camera and so on.

\subsubsection{IDSuEyeInputHandler}\label{sec:IDSuEyeInputHandler}

\begin{lstlisting}
bool IDSuEyeInputHandler::openCams(int left_cam,int right_cam)
{
  hcam_[0] = left_cam;
  is_InitCamera(&hcam_[0], NULL);
  is_SetColorMode(hcam_[0], IS_CM_RGBA8_PACKED);
  is_SetDisplayMode(hcam_[0], IS_SET_DM_DIB);
  is_SetExternalTrigger(hcam_[0], IS_SET_TRIGGER_SOFTWARE);
  // start capture and wait for first image to be in memory
  is_CaptureVideo(hcam_[0], IS_WAIT); 
  switchAutoSensorGain(1);
  switchAutoSensorShutter(1);

  // same for right_cam
  [...]
  
  // add memory to cam
  is_AllocImageMem(hcam_[cam], CAMERA_WIDTH, CAMERA_HEIGHT, CAMERA_DEPTH*CAMERA_CHANNELS, &new_mem_addr, &new_mem_id);
  cam_img_mem_[cam].push_back(std::make_pair(new_mem_addr,new_mem_id));
  is_AddToSequence(hcam_[cam], cam_img_mem_[cam].back().first, cam_img_mem_[cam].back().second);
  is_SetImageMem(hcam_[cam], cam_img_mem_[cam].back().first, cam_img_mem_[cam].back().second);
}
\end{lstlisting}

The \textbf{IDSuEyeInputHandler::openCams()} function starts the communication with the uEye vendor driver and does some camera configurations like setting the color mode, switch auto sensor gain and switch auto sensor shutter. Further on, also memory is added to the cameras.

\begin{lstlisting}
 bool IDSuEyeInputHandler::grabFrame(int cam)
{
  is_LockSeqBuf(hcam_[cam - 1], IS_IGNORE_PARAMETER, last_img_mem);
  memcpy(driver_data, last_img_mem, CAMERA_BUFFER_LENGTH);
  is_UnlockSeqBuf(hcam_[cam - 1], IS_IGNORE_PARAMETER, last_img_mem);

  if (cam == 1)  // no camera flip needed
  { 
    WaitForSingleObject(cameraMutexLeft_, INFINITE);
    memcpy(cameraBufferLeft_, driver_data, CAMERA_BUFFER_LENGTH);
    ReleaseMutex(cameraMutexLeft_);
  } 
  else  // camera flip needed image
  {
    WaitForSingleObject(cameraMutexRight_, INFINITE);
    unsigned char *buffer = cameraBufferRight_;
    char *driver_buffer = driver_data + CAMERA_BUFFER_LENGTH;
    int byte_per_pixel = (CAMERA_CHANNELS * CAMERA_DEPTH) / 8;
    for (int pixel_id = 0; pixel_id < CAMERA_WIDTH * CAMERA_HEIGHT; pixel_id++)
    {
      memcpy(buffer, driver_buffer, byte_per_pixel);
      buffer += byte_per_pixel;
      driver_buffer -= byte_per_pixel;
    }
    ReleaseMutex(cameraMutexRight_);
  }
}
\end{lstlisting}
 Since one of the cameras had to be placed upside down onto the frontplate of the Oculus Rift, that camera image has to be flipped.

\subsubsection{OculusHMD}\label{sec:OculusHMD}

The OculusHMD instance is communicating directly with the Oculus SDK Framework. It is basically doing the device creation here. The details of the SDK's functions can be read in the Oculus Developer Guide \cite{oculusDeveloperGuideV006}.

\begin{lstlisting}
void OculusHMD::calculateFOV()
{
  for (int eye = 0; eye<2; eye++)
  {
     eyeSize_[eye] = ovrHmd_GetFovTextureSize(hmd_, (ovrEyeType)eye,
        hmd_->DefaultEyeFov[eye], 1.0f);
  }
}
\end{lstlisting}

The method is calculating the correct FOV texture size. Since the scene has to be rendered into two 2D textures for the Rift, it is essential that these textures have the exact texture size to guarantee the desired FOV of the HMD hardware.

\begin{lstlisting}
void OculusHMD::configureStereoRendering()
{
  ovrD3D11Config d3d11cfg;
  d3d11cfg.D3D11.Header.API = ovrRenderAPI_D3D11;
  d3d11cfg.D3D11.Header.BackBufferSize = Sizei(hmd_->Resolution.w, hmd_->Resolution.h);
  d3d11cfg.D3D11.Header.Multisample = 1;
  d3d11cfg.D3D11.pDevice = graphicsAPI_->GetDevice();
  d3d11cfg.D3D11.pDeviceContext = graphicsAPI_->GetDeviceContext();
  d3d11cfg.D3D11.pBackBufferRT = graphicsAPI_->rendertargetview_;
  d3d11cfg.D3D11.pSwapChain = graphicsAPI_->swapchain_;

  ovrHmd_ConfigureRendering(hmd_, &d3d11cfg.Config,
  ovrDistortionCap_Chromatic | ovrDistortionCap_Overdrive,
  hmd_->DefaultEyeFov, eyeRenderDesc_))

  useHmdToEyeViewOffset_[0] = eyeRenderDesc_[0].HmdToEyeViewOffset;
  useHmdToEyeViewOffset_[1] = eyeRenderDesc_[1].HmdToEyeViewOffset;
  ovrHmd_GetEyePoses(hmd_, 0, useHmdToEyeViewOffset_, eyeRenderPose_, NULL);
  ovrHmd_AttachToWindow(OculusHMD::instance()->hmd_, graphicsAPI_->window_, NULL, NULL);

  // disable health and security warnings 
  ovrHmd_DismissHSWDisplay(hmd_);
}
\end{lstlisting}
The graphic buffers sizes are configured here for the hardware device and also the swapchain, rendercontext and renderdevice.

\begin{lstlisting}
void OculusHMD::StartFrames()
{
  ovrHmd_BeginFrame(hmd_, 0);
}
\end{lstlisting}
Calls a function from the Oculus SDK which is acting equivalently like the normal BeginScene() method from a render loop.

\begin{lstlisting}
bool OculusHMD::RenderDistortion()
{
  ovrD3D11Texture eyeTexture[2];
  Sizei size;
  size.w = RIFT_RESOLUTION_WIDTH; 
  size.h = RIFT_RESOLUTION_HEIGHT;

  // Stereo Eye Render
  ovrRecti eyeRenderViewport[2];
  eyeRenderViewport[0].Pos = Vector2i(0, 0);
  eyeRenderViewport[0].Size = size;

  eyeTexture[0].D3D11.Header.API = ovrRenderAPI_D3D11;
  eyeTexture[0].D3D11.Header.TextureSize = size;
  eyeTexture[0].D3D11.Header.RenderViewport = eyeRenderViewport[0];
  eyeTexture[0].D3D11.pTexture = graphicsAPI_->renderTextureLeft_->renderTargetTexture_;
  eyeTexture[0].D3D11.pSRView = graphicsAPI_->renderTextureLeft_->GetShaderResourceView();
  // same for eyeRenderViewport[1] with renderTextureRight_
  [...]
  
  ovrHmd_EndFrame(hmd_, eyeRenderPose_, &eyeTexture[0].Texture);
}
\end{lstlisting}
This method's purpose is to supply the Oculus SDK with both of rendered \textbf{RenderTexture}s. During the call to \textbf{ovrHmdEndFrame()} also the SDK barrel distortion is applied to the image as a post-processing effect. 

\begin{lstlisting}
void OculusHMD::trackMotion(float& yaw, float& eyepitch, float& eyeroll)
{
  ovrTrackingState tracking_state = ovrHmd_GetTrackingState(hmd_, ovr_GetTimeInSeconds());

  if (tracking_state.StatusFlags & (ovrStatus_OrientationTracked | ovrStatus_PositionTracked))
  {
    OVR::Posef pose = tracking_state.HeadPose.ThePose;
    pose.Rotation.GetEulerAngles<Axis_Y, Axis_X, Axis_Z>(&yaw, &eyepitch, &eyeroll);

    yaw = RadToDegree(yaw);
    eyepitch = RadToDegree(eyepitch);
    eyeroll = RadToDegree(eyeroll);
  }
}
\end{lstlisting}
Using the Oculus SDK, it is very easy to gather the hardware's current rotation state and save it (radians to degree).

\subsubsection{Shader}\label{sec:Shader}

\begin{lstlisting}
bool Shader::InitializeShader(ID3D11Device* device, HWND hwnd, WCHAR* vsFilename, WCHAR* psFilename, WCHAR* undistShaderFilename)
{
  // Compile all 3 shader programs from file
  result = D3DCompileFromFile(vsFilename, NULL, NULL, "LightVertexShader", "vs_5_0",
    D3D10_SHADER_ENABLE_STRICTNESS, 0, &vertexShaderBuffer, &errorMessage);
  [...] // Fragment shader
  [...] // Undistortion shader

  // Fragment Shader (Texture Mapping, Illumination)
  result = device->CreatePixelShader(pixelShaderBuffer->GetBufferPointer(), 
    pixelShaderBuffer->GetBufferSize(), NULL, &pixelshader_);
  [...] // Fragment Shader (Undistortion)
  [...] // Vertex Shader (Transf, Proj.)

  // 3D Vertices
  polygonLayout[0].SemanticName = "POSITION";
  [...]
  // 2D Texture Coordinates
  polygonLayout[1].SemanticName = "TEXCOORD";
  [...]
  // Normals
  polygonLayout[2].SemanticName = "NORMAL";
  [...]

  // assign the vertex layout.
  result = device->CreateInputLayout(polygonLayout, numElements, vertexShaderBuffer->GetBufferPointer(),
    vertexShaderBuffer->GetBufferSize(), &layout_);

 // set uniforms withing the shader program
  result = device->CreateBuffer(&lightBufferDesc, NULL, &lightBuffer_);
  [...] // matrixBuffer
  [...] // undistortionBuffer
}
\end{lstlisting}
Reads in the shader code for the Vertex-, Fragment- and Undistortion Shader from a file, compiles it and sets the Polygonlayout of the 3D vertices, 2D texture coordinates and normals. Also copys the uniform buffers within the shader programs, which are the matrixBuffer, lightbuffer and undistortionBuffer.

\begin{lstlisting}
void Shader::RenderShader(ID3D11DeviceContext* deviceContext, int indexCount, bool undistort)
{
  // vertex layout
  deviceContext->IASetInputLayout(layout_);

  deviceContext->VSSetShader(vertexshader_, NULL, 0);
  if (undistort)
    deviceContext->PSSetShader(undistortionShader_, NULL, 0);
  else
    deviceContext->PSSetShader(pixelshader_, NULL, 0);

  deviceContext->PSSetSamplers(0, 1, &samplestate_);

  // render
  deviceContext->DrawIndexed(indexCount, 0, 0);
}
\end{lstlisting}
Does the actual graphic rendering and sets the shaders which should be used.

\subsection{Program Configuration}\label{sec:config}
The program has some different configuration possibilities, which are handled by using the following defines within the code:

   \begin{itemize}
      \item\texttt{OpenLabNight\_DEMO} - enables an animated virtual scene which has been shown at the OpenLabNight of the ICG institute
      \item\texttt{AR\_HMD\_ENABLED} - configure if a HMD is connected to the computer
      \item\texttt{HMD\_DISTORTION} - configure if the barrel distortion from the Oculus SDK is desired
      \item \texttt{SHOW\_FPS} - enable Frames-Per-Second visuals
      \item \texttt{FULL\_SCREEN} - start the program in full screen or window mode
      \item\texttt{RIFT\_RESOLUTION\_WIDTH} - set the HMDs full display resolution
      \item\texttt{RIFT\_RESOLUTION\_HEIGHT} - set the HMDs full display resolution
      \item\texttt{CAM1} - camera 1 device ID
      \item\texttt{CAM2} - camera 2 device ID
       \item\texttt{CAMERA\_CHANNELS} - set camera channels
       \item\texttt{CAMERA\_WIDTH } - set camera width
       \item\texttt{CAMERA\_HEIGHT} - set camera height
       \item\texttt{CAMERA\_DEPTH} - set camera bit depth
       \item\texttt{CAMERA\_BUFFER\_LENGTH} - compute camera memory in bytes                                                                                                                                                                       
      \item\texttt{SCREEN\_DEPTH} - configure the virtual far plane
      \item\texttt{SCREEN\_NEAR} - configure the virtual near plane
   \end{itemize}

\subsection{Libraries}\label{sec:library}

   \begin{itemize}
      \item Windows SDK Win8.1
      \item DirectX ToolKit
      \item Oculus SDK 0.4.4
      \item IDS uEye Driver
      \item OCamCalib Matlab Toolbox
   \end{itemize}

\section{Conclusion}\label{sec:conclusion}

Since all the other AR hardware devices like the Google Glass and Microsoft's Hololens are limited in case of augmentation space due to the screen limitations, the immersion with the Oculus Rift  is very intense compared to them. Through the stereoscopic view, the holograms appear like they are really in the real world. And since the HMD hardware completely masks out reality, it is possible to augment the whole screen space for both eyes. This yields the huge AR immersion.

Developing a new engine from scratch with DirectX was a good decision instead of using an already existing graphics engine, since it gives full control over what the engine does and how it does it because it is all C++ code. With that, a huge degree of freedom is achieved since the development takes place at the "hardware's metal". This decision was also made for the sake of extensibility, since it was already clear right from the beginning that the AR engine with the Oculus Rift will be extended with more other features and algorithms.

Previously, the undistortion of the camera images were computed using OpenCV library, but the high distortion could not be handled on the CPU as fast due to lower fps. That's why the undistortion is computed on a fragment shader within the engine. The performance gain was very satisfying, without any fps dropdowns.

Since the Oculus Rift integration is very clean encapsulated in the code, it should be very straight forward to also integrate other HMDs if that should be desired in the future.

\begin{figure}[H]
  \centering
  \includegraphics[width=0.9\textwidth]{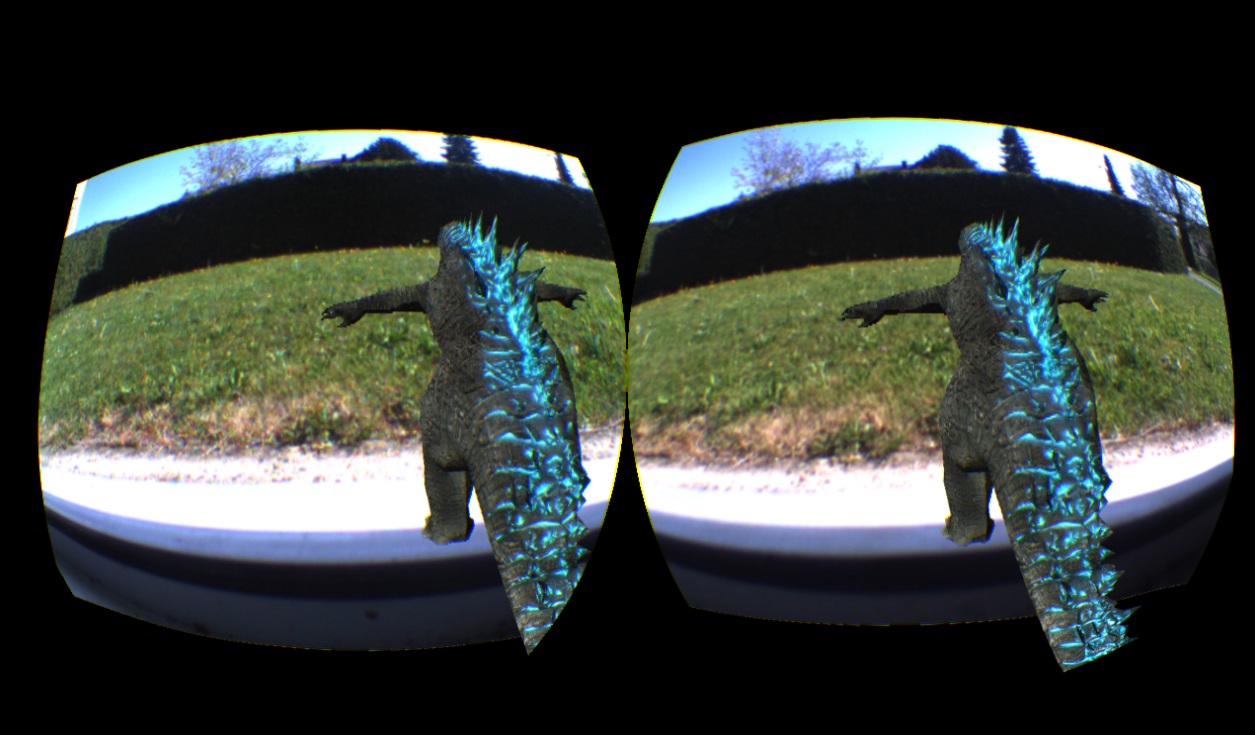}
  \caption{Results - Godzilla}
  \label{fig:godzilla}
\end{figure}

\begin{figure}[H]
  \centering
  \includegraphics[width=1.0\textwidth]{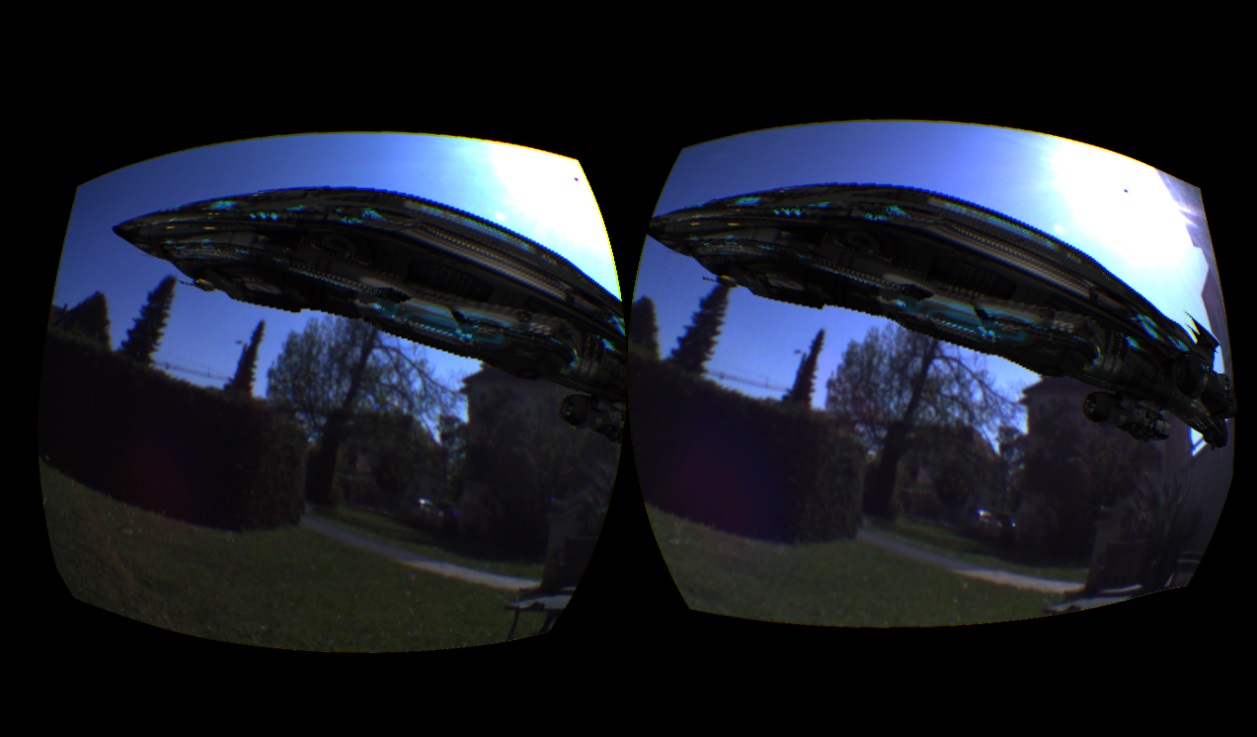}
  \caption{Results - Starship}
  \label{fig:spaceship}
\end{figure}

\begin{figure}[H]
  \centering
  \includegraphics[width=1.0\textwidth]{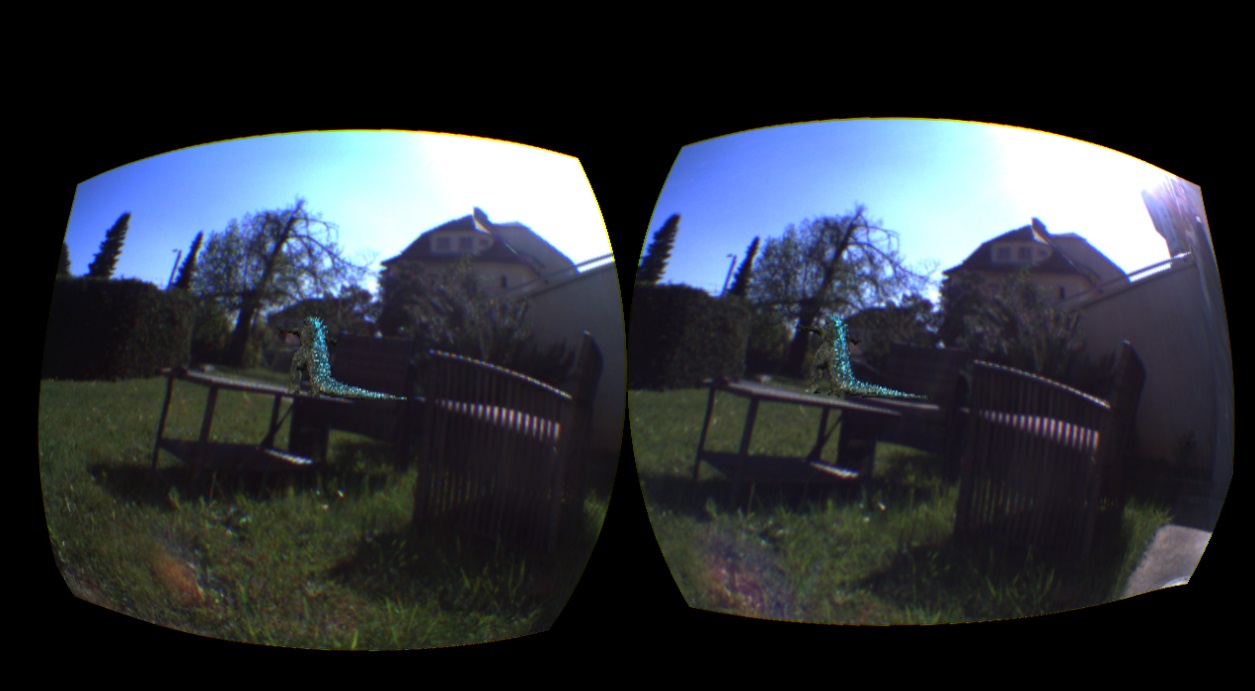}
  \caption{Results - Godzilla Tiny}
  \label{fig:dino}
\end{figure}

\begin{figure}[H]
  \centering
  \includegraphics[width=0.9\textwidth]{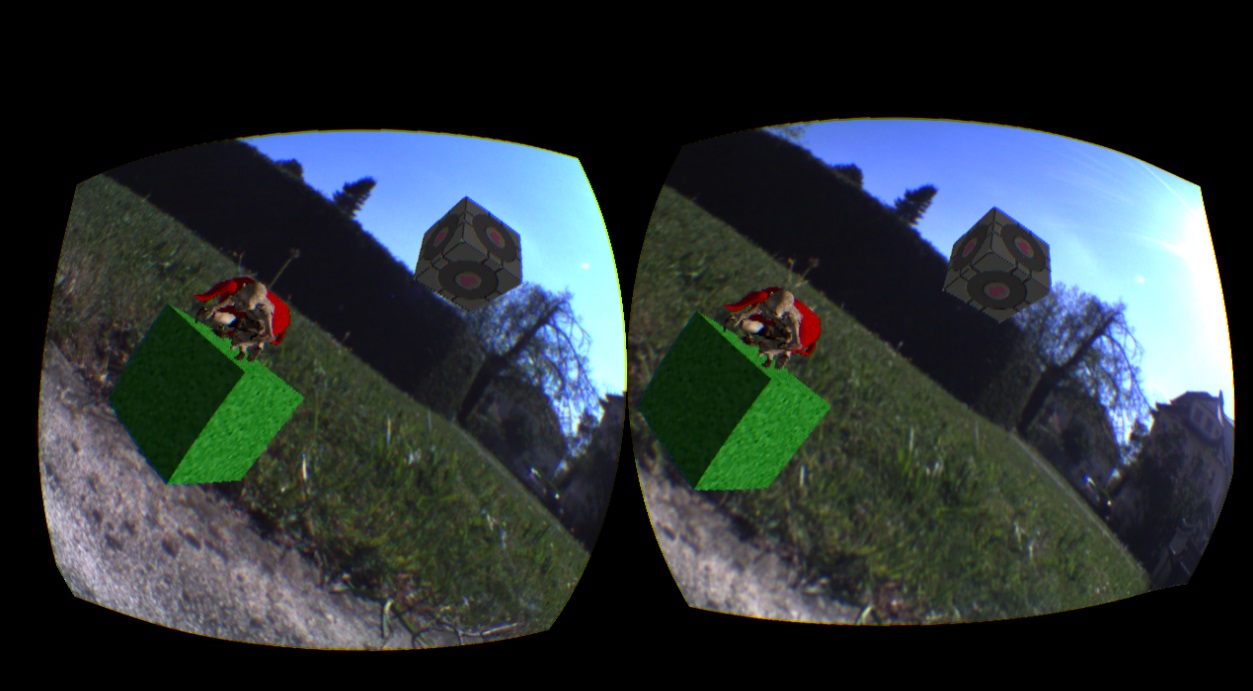}
  \caption{Results - Multiple Holograms}
  \label{fig:multi_objects}
\end{figure}

However, there is one bottleneck in case of extensibility and that is the basic camera capturing. Since both of the cameras are USB 2.0 cameras, it was not possible to stream both cameras over the same USB HUB using OpenCV, which would establish an easy way of camera streaming. The bus bandwith with USB 2.0 cameras simultaneously was not sufficient to handle the huge data loads. That is why it was necessary to write the camera streaming manually again "on the metal", using the IDS uEye drivers. If in future other cameras are desired, it would be a good decision to consider one from the same manufactorer since they will probably rely on the same hardware driver. Figure \ref{fig:godzilla}, \ref{fig:spaceship}, \ref{fig:dino} and \ref{fig:multi_objects} show some final results from the AR stereo engine. The model files and textures shown in the sample results are taken from web sources like TF3DM under educational usage purposes.

\section{Future Work}\label{sec:future}

Since the project's results are very satisfying, it is already clear how the engine will be extended and evolve in future.
One of them is an integration of the \textbf{Large-Scale Direct Monocular Simultaneous Localization and Mapping (LSD-SLAM)} algorithm \cite{engel2014lsd}. LSD-SLAM outclasses the common augmentation method of pattern placements in the real world to recognize the virtual viewpoint. The algorithm keeps track of the camera translation and rotation and computes intern a 3D pointcloud of an unknown environment, which makes it possible to really move in both worlds without any pattern placements.

The second project will be an integration of a realtime \textbf{Hand Tracking} algorithm \cite{oberweger2015training} for hand pose estimation. This will make it possible to directly interact with the 3D holograms using hand gestures. Also a hand segmentation could be done, which will make it more realistic when interacting with the virtual objects. We are already assigned to both future projects.

Another one is to design a 3D printed hardware attachment to put on 2 cameras at will on the frontplate of the Oculus Rift. This makes it possible to share the AR attachemenet over the world with other people.

\clearpage
\bibliography{library}
\bibliographystyle{ieeetr}
\appendix

\end{document}